\documentclass[12pt]{article}
\linethickness{0.4pt}
\titlepage
\newcommand{\be}{\begin{equation}}
\newcommand{\ee}{\end{equation}}
\newcommand{\bea}{\begin{eqnarray}}
\newcommand{\eea}{\end{eqnarray}}
\author{O. Diego\footnote{EMail: odiego@apiaxxi.es
Apia XXI, Departamento de Topograf\'\i a, 
Luis Mart\'\i nez 21, 39005 Santander, Spain} \footnote{EMail:miodiego@hotmail.com}}
\title{Toward the unification of the postulates of Quantum Mechanics}

\begin{document}
\maketitle
\abstract{In this paper we are going to introduce a new dynamical postulate in Quantum Mechanics. This new principle 
is defined using path integrals over the set of normalized wave functions. We 
will show in a qualitative way that this postulate is equivalent to the Schr\"odinger equation and to 
the measurement postulates. Then we propose a new set of fundamental postulates for Quantum Mechanics. 
In this approach to Quantum Mechanics we separate the fundamental postulates and the physical 
interpretation. The fundamental postulates are abstract mathematical principles. Their physical 
interpretation depend on the physical system under study. These postulates do not use the 
concepts of measurement device and observer.}
\newpage

{\bf 1. Introduction}

In Quantum Mechanics measurement devices play a very important role. 
They are essential for the physical interpretation of Quantum Mechanics. 
In fact, isolated physical systems without measurement devices have no physical meaning. 
In Quantum Mechanics the word probability means the probability of have some result 
if some measurement is performed.

Moreover, physical systems are defined by solutions of the Schr\"o\-ding\-er equation whereas 
measurement devices are governed by their own postulates. There are two basic mathematical 
operations in Quantum Mechanics: the Schr\"o\-ding\-er equation and the scalar product. 
The Schr\"o\-ding\-er equation represents the reversible smooth temporal evolution 
of isolated physical systems. The scalar product is interpreted as the irreversible sudden collapse 
of physical states when a measurement is performed.

Measurement devices are not mathematical abstract objects but they are real physical objects. 
Hence, why should exist different physical laws for measurement devices and physical 
systems?. The postulates of orthodox Quantum Mechanics are firmly anchored in the experimental 
results of the last century. But it is possible that these postulates are not fundamental. 
They may come from more fundamental postulates that 
do not distinguish between physical systems and measurement devices.

The main purpose of this paper is to look for new postulates that unify the Schr\"o\-ding\-er 
equation and the measurement postulates. The main assumption of this paper is that physical 
systems and measurement devices are governed by the same physical laws.

First of all we must show if the Schr\"o\-ding\-er equation and the measurement postulates are 
independent postulates. Measurement devices are sets of atoms. Hence we can write the full 
Schr\"o\-ding\-er equation of physical systems and measurement devices.
¿Is the full Schr\"o\-ding\-er equation equivalent to the measurement postulates?

The answer to this question seem to be negative. But there is not a clear difference 
between physical systems and measurement devices. For instance, 
let us consider the physical system $S$ that is observed by the measurement device $M_{1}$. 
Now, let us consider the measurement device $M_{2}$ that observe both $S$ and $M_{1}$. 
Then it is possible to apply the Schr\"o\-ding\-er equation to $S$ only. Or it is also possible 
to apply the Schr\"o\-ding\-er equation to $S$ and $M_{1}$. But it seems that we cannot apply the
full Schr\"o\-ding\-er equation to $S$, $M_{1}$ and $M_{2}$. In other words sometimes measurement devices 
can be regarded as part of the physical system under study. But something must play the 
role of measurement device. In orthodox Quantum Mechanics physical systems without measurement
devices have no sense\cite{1}. 

Actually, if we try to describe the behaviour of measurement devices with the Schr\"o\-ding\-er 
equation alone, 
we will find three important problems.

First of all there is the problem of the classical limit of the Schr\"o\-ding\-er equation.
Measurement devices are classical objects. Therefore, if the Schr\"o\-ding\-er equation must 
describe measurement devices then it must have a reasonable classical limit. 
The theorem of Ehren\-fest\cite{2} suggests that mean values of observables are 
the corresponding classical variables in the classical limit.
But there are physical states that yield mean values without classical limit. 
For instance, the mean value of the electric field operator is zero in states with a fixed number of photons. 
On the other hand the Schr\"o\-ding\-er equation is linear but the superposition 
principle does not hold in the classical limit. Hence, there are solutions 
of the Schr\"o\-ding\-er equation that cannot define physical states in the classical limit.
Hence the Schr\"o\-ding\-er equation have a reasonable classical limit if truncate the 
set of its solutions. 

The second problem is the collapse of the wave function.
There are solutions of the full Schr\"o\-ding\-er equation that do not collapse.
For instance, let us consider the system S and the measurement devices ${M_{i}}$. 
Let us call H to the full Hamiltonian of S, $\{M_{i}\}$ and their interactions. 
The eigenfunctions of the full hamiltonian H do not evolve and they cannot collapse.
Hence we cannot give an orthodox interpretation to some solutions of the full 
Schr\"o\-ding\-er equation.

The third problem is that the Schr\"o\-ding\-er equation is deterministic. Hence the 
Schr\"o\-ding\-er equation alone cannot yield the probabilistic interpretation of 
Quantum Mechanics.

But the Schr\"o\-ding\-er equation alone almost can yield the measurement postulates. For instance, in 
an old paper\cite{6} Mott studied the full Schr\"o\-ding\-er equation of a particle interacting with 
a Wilson chamber. He found solutions that can be interpreted as tracks leaving by the particle. 
But the Schr\"o\-ding\-er equation cannot explain why these are the only solutions observed in nature. 
Linear combination of these solutions are also solutions but they cannot represent any physical process. 
On the other hand the Schr\"o\-ding\-er equation has been used in the study of macroscopic systems. 
But in these cases there are solutions of the Schr\"o\-ding\-er equation that are non physical. 
These solutions are forbidden at hand. Hence we only need some kind of principle that truncates 
the set of physical states.

Despite the successes of orthodox Quantum Mechanics there has been several attempts 
to introduce new fundamental postulates.

One of the first attempt was the pilot wave theory of de Broglie-Bohm\cite{3}. 
In this theory particles are governed by deterministic physical laws.
The initial conditions are not known exactly. This is the origin of the probabilistic 
behaviour of Quantum Mechanics.

The main problem of this theory is that it is unnatural that we cannot know exactly the 
initial conditions if the fundamental laws are deterministic. Actually this theory belongs 
to the set of theories called hidden variables theories. According to these theories there 
is some kind of deterministic laws underlying the probabilistic nature of Quantum Mechanics. 
Recently Gerard`t Hooft
has proposed a new theory of this kind\cite{Hooft}. 

But the main problem is not the probabilistic nature of the postulates of orthodox Quantum Mechanics. 
The main problem is the difference between physical systems and measurement devices. 
In hidden variables models we must distinguish between physical states that evolve 
in a deterministic way and initial conditions that are not given by an exact state. 
Initial conditions must be given by some kind of probability. This is very important because 
if the fundamental laws are deterministic and initial conditions are fixed then all is determined. 
But we need to explain the origin of probability distributions at the initial time. 
If we said that the observer do not know the exact initial state but it has an incomplete 
information then observers will play an essential role in these theories.  

More recently there has been several attempts\cite{4} to show that the Schr\"o\-ding\-er 
equation of open systems can yield the orthodox postulates.

In this approach one start from the full Schr\"o\-ding\-er equation of the physical system,
the measurement devices and the environment.
The environment is described by an statistical mixtures of states.
The main problem of this approach is that it does not work if the environment is 
described by a pure state. But the Schr\"o\-ding\-er equation alone cannot yield 
restrictions over the set of physical states.

The environment is given by a very large number of particles. Therefore in practical calculations 
it is very natural to make the assumption that it is defined by a statistical mixtures of states. 
But the problem is that this assumption does not arise from the Schr\"o\-ding\-er equation 
or other fundamental principle. It arise from our empirical experience. If this assumption 
is transformed into a fundamental principle then physical systems and the environment 
obey different physical laws.
 
The approach introduced by Ghirardi, Rimini and Weber\cite{5} has attracted some attention.
In this approach all physical systems, including measurement devices, evolve according to 
the Schr\"o\-ding\-er equation. But sometimes there is an spontaneous collapse. These collapses 
is not consequence of measurements processes 
but it holds with some probability for all physical systems. The problem is that energy is 
not conserved because these collapses do not arise from interactions with measurement devices. 

Let us consider the lineal combination 
\be
\Psi = \Psi_{1}+\Psi_{2}.
\label{eq:1.1}
\ee
The Schr\"o\-ding\-er equation cannot transform ($\ref{eq:1.1}$) into the temporal evolution 
of $\Psi_{1}$ or $\Psi_{2}$. But the spontaneous collapse can do it with some probability. 
This probability is very small for microscopic system but it is large for macroscopic systems.
Hence, decoherence holds for macroscopic systems in this formalism.

Let 
\bea
\left\langle \Psi_{1} \left| H \right| \Psi_{1} \right\rangle & = & E_{1} \nonumber \\
\left\langle \Psi_{2} \left| H \right| \Psi_{2} \right\rangle  & = & E_{2}.
\label{eq:1.2}
\eea
If $E_{1}$ and $E_{2}$ are different, then the transformation of the lineal 
combination ($\ref{eq:1.1}$) do not conserve the energy.

Therefore the spontaneous collapse that transform the linear combination ($\ref{eq:1.1}$) 
into one of its components can violate the conservation energy law. We can consider only 
that spontaneous collapses of the wave function transform the linear combination ($\ref{eq:1.1}$)
into the statistical mixtures of its components. In this case energy, defined as a mean value of $E_{1}$ 
and $E_{2}$, is not conserved. But the violation of the energy conservation law is very small. 
Actually it is consistent with physical experiments. 

But now we have the problem of the interpretation of the statistical mixture of states. 
If we make the usual interpretation, the statistical mixture only means that the state is a 
pure state $\Psi_{1}$ or $\Psi_{2}$ but we do not know it. Therefore in this case energy is not conserved. 

All these approaches and orthodox Quantum Mechanics have in common that they introduce two 
different physical laws: The Schr\"o\-ding\-er equation and some other probabilistic postulate.
There are two possibilities. It is possible that there are two different objects which 
obey different physical laws.
Or it is possible that the two different physical laws work at different times.
In this paper we are going to introduce a new dynamical principle 
that unify the Schr\"o\-ding\-er equation and the measurement postulates.

\newpage

{\bf 2.} 

The main purpose of this paper is to look for a new set of postulates without the 
concept of measurement device. The founders of Quantum Mechanics have established 
that physical interpretation must use the concepts of Classical Mechanics.
But we cannot define physical states of quantum systems with classical variables.

Measurement devices are classical objects. Measurement processes yield well defined 
classical variables. In orthodox Quantum Mechanics classical concepts are introduced 
through measurement devices.

We are going to look for new fundamental postulates without the concept of measurement 
device. Hence physical interpretations cannot enter into the fundamental postulates.

There are not contradictions between orthodox Quantum Mechanics and experimental results.
Why should we reject the orthodox physical interpretation of Quantum Mechanics?.

The postulates of orthodox Quantum Mechanics can only be applied to physical systems 
interacting with measurement devices. Hence, isolated microscopic systems have no sense.
This is not important in practical applications. There are always measurement devices 
in every real experiment.

But from a fundamental point of view we have the following problem. There is nothing beyond 
measurement processes in the orthodox interpretation of Quantum Mechanics.

Is there a real world beyond measurement processes?.
This is a philosophical question and its answer is a matter of opinion.
Some founders of Quantum Mechanics believed that Physic must describe only experiments.
But today most physics believe that there is a real world that we try to understand.
Superstring theories or Quantum Gravity theories have no sense if we do not believe 
in the existence of a real world beyond the experiment facts.

In this paper we are going to start from a realistic philosophical point of view.
Hence we must look for new postulates that can work for all physical systems. We must faced 
with two essential problems. Measurement devices enter into the fundamental postulates 
through the physical interpretation of the mathematical objects defined in orthodox Quantum Mechanics. 
Hence we must separate the fundamental laws and the physical interpretation. The fundamental laws will 
be mathematical abstract principles. 
The second problem is the unification of the smooth reversible evolution of the Schr\"o\-ding\-er 
equation and the sudden irreversible collapse measurement processes into a single dynamical principle. 
Hence the new principle cannot be given by a temporal evolution equation and time will not 
be a fundamental concept.

We are going to introduce two postulates. A cinematic principle that tell us how to describe
physical systems and a dynamical principle.

The first postulate is that physical states of N particles are defined by wave functions:
\be
\Psi(t;{\bf r}_{1},\cdots,{\bf r}_{N}).
\label{eq:2.1}
\ee
The orthodox physical interpretation of wave functions uses the concept of measurement device.
Hence wave functions have only sense for system interacting with measurement devices.

In our approach wave functions are defined for all physical systems. But they are abstract 
mathematical objects. Only in some cases we can give a physical interpretation of wave functions.
Hence the parameters $\{\bf{r}_{i}\}$ are mathematical abstract objects.
In some cases these parameters can be interpreted as coordinates in some space.

In orthodox Quantum Me\-cha\-nics isolated mi\-cros\-co\-pic systems have no sense.
In this new approach these systems have sense.
But they are described by wave functions that can not be interpreted in the usual way.

Now we are going to introduce the dynamical postulate.
We are going to unify the Schr\"o\-ding\-er equation and the measurement postulates into a single 
postulate.
This postulate cannot be a partial differential equation.
Sometimes for a fixed initial wave functions there must be several final wave functions.

The second postulate will be the definition of the correlator 
\be
\left\langle \Psi_{1}(t_{1}) | \Psi_{2}(t_{2}) \right\rangle.
\label{eq:2.2}
\ee
In some cases the correlator ($\ref{eq:2.2}$) can be interpreted as a transition amplitude.
But the correlator ($\ref{eq:2.2}$) is an abstract mathematical object at the fundamental level.

Let us define ($\ref{eq:2.2}$) using the path integral approach
\be
\left\langle \Psi_{1}(t_{1}) | \Psi_{2}(t_{2}) \right\rangle = \int D \Omega [\Psi] \, \, \, Weight[\Psi],
\label{eq:2.3}
\ee
where the measure 
\be
\int D \Omega [\Psi]
\label{eq:2.4}
\ee
represents the sum over all wave functions 
\be
\Psi(t;{\bf r}_{1},\cdots,{\bf r}_{N}) ,
\label{eq:2.5}
\ee
with boundary conditions 
\bea
\Psi(t_{1}) & = & \Psi_{1}(t_{1}) \nonumber \\
\Psi(t_{2}) & = & \Psi_{2}(t_{2}). 
\label{eq:2.6}
\eea
The wave functions $\Psi_{1}$ and $\Psi_{2}$
are not arbitrary but they must verify
\bea
\int \prod_{i=1}^{N} d^{3} {\bf r}_{i} \Psi_{1}^{*} 
\hat{H} \Psi_{1} & = & 
\int \prod_{i=1}^{N} d^{3} {\bf r}_{i} \Psi_{2}^{*} 
\hat{H} \Psi_{2} \nonumber \\
\int \prod_{i=1}^{N} d^{3} {\bf r}_{i} \Psi_{1}^{*} 
\hat{\bf{P}} \Psi_{1} & = & 
\int \prod_{i=1}^{N} d^{3} {\bf r}_{i} \Psi_{2} 
\hat{\bf{P}} \Psi_{2} \nonumber \\
\int \prod_{i=1}^{N} d^{3} {\bf r}_{i} \Psi_{1}^{*} 
\hat{\bf{L}} \Psi_{1} & = & 
\int \prod_{i=1}^{N} d^{3} {\bf r}_{i} \Psi_{2}^{*} 
\hat{\bf{L}} \Psi_{2}, 
\label{eq:2.7}
\eea
where $\hat{H}$, $\hat{\bf P}$ and $\hat{\bf L}$
are the total hamiltonian, momentum and angular momentum operators.
In this formalism ($\ref{eq:2.7}$) are mathematical abstract conditions.
In some particular cases ($\ref{eq:2.7}$) will be interpreted as the conservation of energy, 
momentum and angular momentum.

The measure ($\ref{eq:2.4}$) is defined only over normalized wave functions. In other words
\be
\int \prod_{i=1}^{N} d^{3} {\bf r}_{i}  
\left| \Psi(t;{\bf r}_{1} \cdots, {\bf r}_{N}) \right|^{2} = 1 \, \, \, \, \forall t. 
\label{eq:2.8}
\ee

Let us define the measure ($\ref{eq:2.4}$) by:
\be
\int D \Omega [\Psi] = \int \prod_{t} \prod_{i=1}^{N} \prod_{{\bf r}_{i}} \left\{
\frac{d \Psi(t;{\bf r}_{1} \cdots, {\bf r}_{N}) d \Psi(t;{\bf r}_{1} \cdots, {\bf r}_{N})}
{\left|\Psi(t;{\bf r}_{1} \cdots, {\bf r}_{N}) \right|^{4}} \right\}.
\label{eq:2.9}
\ee

This is a symbolic expression. In order to define ($\ref{eq:2.9}$) in a rigorous way we must discretize 
the continuum variables $ ( t; \{ {\bf r}_{i} \} )$.
The path integral becomes a multiple Riemann integral.
Then we perform the calculations and we perform the continuum limit.
This is analogous to the definition of Path Integrals in Quantum Field Theories.

In Quantum Mechanics path integrals can be defined rigorously.
But in Quantum Field Theory the continuum limit can depend on the discretization method.

Discretized path integrals are analogous to partition functions in Statistical Mechanics.
Near second order phase transitions the physical behaviour does not depend on the discretization.
Therefore the path integral ($\ref{eq:2.3}$) has sense if the corresponding statistical model 
has second order phase transitions.

We are not going to perform explicit calculation but we are going to accept that the path 
integral ($\ref{eq:2.3}$) has sense and we are going to study the discrete version 
of ($\ref{eq:2.3}$).

Let us define the weight of ($\ref{eq:2.3}$) by
\be
Weight[\Psi] = \exp{\left[\frac{i}{\alpha \hbar} \int_{t_{1}}^{t_{2}} \int \prod_{i=1}^{N} 
d^{3} {\bf r}_{i} \Psi \hat{O} \Psi \right]},
\label{eq:2.10}
\ee
where the operator $\hat{O}$ is given by
\be
\hat{O}= - i \hbar \frac{\partial}{\partial t} - \hat{H}
\label{eq:2.11}
\ee
and $\hat{H}$ is the hamiltonian. Let us remark that we have introduced a new adimensional 
constant $\alpha$.

We have defined the mathematical objects ($\ref{eq:2.2}$). We have called them correlators. 
But they are mathematical abstract objects. They are not probability amplitudes or S-matrix elements 
or the usual correlators of Quantum Mechanics. Given a physical systems we must compute all correlators. 
Then we can make physical interpretations compatible with these results. 

The Schr\"o\-ding\-er equation works very well for all non relativistic physical
systems: microscopic and macroscopic. But for macroscopic systems there are 
solutions of the Schr\"o\-ding\-er equation that are non physical. Hence we need to introduce some 
kind of truncation over the set of states. There are two possibilities: a cinematic principle or
a dynamic principle. 

A cinematic principle truncates the set of states at the initial time. But the cinematic principle must work 
only for macroscopic systems. But it is difficult to define the concept of 
macroscopic system without taking into account its temporal evolution. Macroscopic systems are not just
systems with a very large number of particles. The particles must interact between them. 

Therefore it seems more natural to introduce a dynamic principle that truncates the set of solutions of the 
Schr\"o\-ding\-er equation at every time. In some alternatives to orthodox Quantum Mechanics 
there are sums over solutions of the Schr\"o\-ding\-er equation. For instance, in the open systems
approach\cite{4} we sum over the set of possible states of the environment. In other words we must 
sum over the set of solutions of the full Schr\"o\-ding\-er equation. The contribution to the sums of
some wave functions are very small. We can eliminate these wave functions. Therefore sums over wave functions
can truncate the set of solutions. But in this approach we must 
sum over wave functions because observers cannot know the exact state of the environment.

We need to have sums over wave functions because we need to compare wave functions and in some cases 
we need to eliminate some solution of the Schr\"o\-ding\-er equation. But at the same
time we do not want that observers enter into the postulates of the theory. Therefore sums over wave
functions must be introduce into the fundamental postulates. Hence we define the 
correlators ($\ref{eq:2.2}$) as sums over wave functions. 

In Quantum Field Theory the second quantization of the Schr\"o\-ding\-er equation yields a path 
integral analogous to ($\ref{eq:2.3}$). Hence the limit 
\be
\alpha \longrightarrow 0 
\label{eq:2.12}
\ee
is analogous to the classical limit.
If $\alpha $ is small the leading contributions to the weight ($\ref{eq:2.10}$) are given 
by solutions of the Schr\"o\-ding\-er equation 
\be
\hat{O} \Psi = \left( -i \hbar \frac{\partial}{\partial t} - \hat{H} \right) \Psi = 0.
\label{eq:2.13}
\ee
The normalization condition ($\ref{eq:2.8}$) is very natural in this formalism. Correlators are defined by sums
over all wave functions. For solutions of the Schr\"o\-ding\-er equation the weight ($\ref{eq:2.10}$)
is just one. But for no solutions of the Schr\"o\-ding\-er equation with norm very small the weight 
is near one. We do not want that arbitrary wave functions with very small norms have roles similar 
to solutions of the Schr\"o\-ding\-er equation. Hence we restrict the path integral to normalized 
wave functions. This condition ($\ref{eq:2.8}$) will play a very important role in the next section.

In the next section we are going to study in a qualitative way the discrete version 
of the measure ($\ref{eq:2.9}$).
We will study the mathematical object ($\ref{eq:2.3}$) for macroscopic and for microscopic systems.
We will show that ($\ref{eq:2.3}$) is equivalent to the Schr\"o\-ding\-er equation for microscopic systems.
Whereas it is equivalent to the Schr\"o\-ding\-er equation restricted to some subset of the set of 
states for macroscopic systems.

In section four, we will study in a qualitative way the interaction between a microscopic system with 
a measurement device.
The results of this section are analogous to the measurement postulates.

In other words, in section three and four, we will show qualitative arguments that suggest that 
the new dynamical postulate ($\ref{eq:2.3}$) can unify the Schr\"o\-ding\-er equation and the 
measurement postulates.

The last section will be the conclusions.

\newpage

{\bf 3.}

The new dynamical principle introduced in the above section must be equivalent to the Schr\"o\-ding\-er equation 
and to the measurement postulates.
In orthodox Quantum Mechanics the Schr\"o\-ding\-er equation is also used to study macroscopic systems.
Macroscopic solids are described by wave functions that are solutions of the Schr\"o\-ding\-er equation.
But only a subset of solutions can define physical states. Physical states must describe ions that
are oscillating around fixed equilibrium positions.

Let us call local wave functions $\Psi_{L}$ 
to those wave functions whose range is a small subset of the configuration space 
$R^{3N}$. Hence macroscopic systems must be defined by local wave functions.
Spreading wave functions are called non local $\Psi_{NL}$.

The problem in orthodox Quantum Mechanics is that the Schr\"o\-ding\-er equation cannot transform a non local
into a local wave function in a small interval of time.
Therefore the Schr\"o\-ding\-er equation alone cannot describe the physical behaviour of macroscopic systems.
We need also some truncation over the set of physical states.

In orthodox Quantum Mechanics we truncate the set of physical states at hand. 
We are going to show that this truncation arises from the new dynamical principle ($\ref{eq:2.3}$).
But macroscopic systems are described by some subset of solutions of the Schr\"o\-ding\-er 
equation.
Therefore the new dynamical principle must be equivalent to the Schr\"o\-ding\-er equation for macroscopic 
systems also.

We have shown that if $\alpha $ is small then the leading contributions to the weight ($\ref{eq:2.10}$) 
are given by solutions of the Schr\"o\-ding\-er equation. We are going to show now that the leading 
contributions to the measure ($\ref{eq:2.4}$) are given by local wave functions.
 
Let us consider a physical system of 
one dimension with one particle.
This system is defined by wave functions like $\Psi(t,x)$.
Let us discretize the continuum variable $x$
\be
x_{n}= a n \, \, \, \, 	n=1,\cdots,M.
\label{eq:3.3}
\ee
In this case the normalization condition ($\ref{eq:2.8}$) becomes
\be
\sum_{n=1}^{M} a \left| \Psi(t,x_{n}) \right|^{2} = 1.
\label{eq:3.4}
\ee
The measure ($\ref{eq:2.9}$) is defined by
\be
\prod_{t} \prod_{n=1}^{M} \frac{d \Psi(t,x_{n}) d \Psi^{*} 
(t,x_{n})}{\left| \Psi(t,x_{n})\right|^{4}}.
\label{eq:3.5}
\ee
Let us consider the homogeneous wave function $\Psi_{H}$ . This function verify the following condition
\be
\left| \Psi_{H}(t,x_{n}) \right| = A \, \, \forall n.
\label{eq:3.6}
\ee
Hence the constant $A$ must be
\be
A = \sqrt{\frac{1}{a M}}.
\label{eq:3.7}
\ee
Therefore the contribution of $\Psi_{H}$ to the measure ($\ref{eq:3.5}$) is proportional to
\be
\prod_{t} \left( a M \right)^{2 M}.
\label{eq:3.8}
\ee
The wave function $\Psi_{H}$ is non local.

Now let us consider the inhomogeneous wave function $\Psi_{I}$ defined by
\bea
\left| \Psi_{I} ( t, x_{1} ) \right| & = & B \nonumber \\
\left| \Psi_{I} ( t, x_{n} ) \right| & = & A \, \, \forall n \neq 1.
\label{eq:3.9}
\eea
Then $A$ must be
\be
A = \sqrt{\frac{1 - a B^{2}}{a ( M-1)}}.
\label{eq:3.10}
\ee
In the continuum limit $M$ goes to infinity and $a$ goes to zero.
The number $B$ is very large such that $aB^{2}$ is near 1.
Hence $\Psi_{I}$ represent a Dirac delta function around the point $x_{1}$.
Therefore $\Psi_{I}$ is a local wave function.

The contribution of the wave function $\Psi_{I}$ to the measure ($\ref{eq:3.5}$) 
is proportional to
\be
\prod_{t} \left\{ B^{2} \frac{\left(1 - a B^{2}\right)^{M-1}}
{\left(a (M-1)\right)^{M-1}}\right\}^{-2}.
\label{eq:3.11}
\ee

The relation between contributions ($\ref{eq:3.8}$) and ($\ref{eq:3.11}$) is
\be
\prod_{t} \left[ \frac{B^{2} \left( 1 - a B^{2}\right)^{M-1} \left(a M \right)^{M}}
{ \left(a (M-1)\right)^{M-1}}\right]^{2}.
\label{eq:3.12}
\ee
In the continuum limit
\bea
M & \longrightarrow & \infty \\
a & \longrightarrow & 0 \nonumber \\
B^{2} & \longrightarrow & \infty \nonumber \\
0& < a B^{2} < & 1 .
\label{eq:3.13}
\eea
Hence ($\ref{eq:3.12}$) becomes
\be
\prod_{t} \left[ M \left( a B^{2} \right) \left( 1 - a B^{2} \right)^{M-1}
\left( \frac{M}{M-1}\right)^{M-1}\right]^{2}.
\label{eq:3.14}
\ee
If $M$ is very large then
\be
\left ( \frac{M}{M-1} \right )^{M-1} = \left( 1 + \frac{1}{M-1} \right)^{M-1} \approx e.
\label{eq:3.15}
\ee
Hence in the continuum limit ($\ref{eq:3.14}$) becomes proportional to
\be
\prod_{t} \left[ M \left(\ 1 - a B^{2} \right)^{M-1}\right]^{2}.
\label{eq:3.16}
\ee
In the continuum limit $aB^{2}$ is finite and
\be
0 < aB^{2} < 1.
\label{eq:3.17}
\ee
Then, in the continuum limit,
\be
\left[ M \left(\ 1 - a B^{2} \right)^{M-1} \right]^{2} << 1.
\label{eq:3.18}
\ee
Therefore the contribution to the measure of the inhomogeneous wave function $\Psi_{I}$ 
is greater than the contribution of the homogeneous wave function $\Psi_{H}$. In general wave 
functions with very small amplitude in large regions of space yield the leading contribution to the 
measure because the factor
\be
\frac{1}{\left| \Psi(t,x_{n})\right|^{4}}.
\label{eq:3.5bis}
\ee
But wave functions cannot be identically zero because the normalization condition ($\ref{eq:3.4}$). 
Hence the leading contributions to the measure ($\ref{eq:2.9}$) are given by local wave functions.
Now we are going to study more carefully the contribution to the full path integral ($\ref{eq:2.3}$).

We are going to consider non local wave functions $\Psi^{H}_{S}$ defined over all space 
that are solution of the Schr\"o\-ding\-er equation
and local wave functions $\Psi^{I}_{NS}$ that are not solutions of the Schr\"o\-ding\-er equation.
The leading contribution to the weight are given the solutions of the Schr\"o\-ding\-er equation, 
like $\Psi^{H}_{S}$, but 
the leading contribution to the measure are given by local wave functions like $\Psi^{I}_{NS}$. 

Let us consider the set of homogeneous wave functions
\be
\Psi = \Psi_{S}^{H} + \Psi_{H}^{\prime}.
\label{eq:3.19}
\ee
The wave function $\Psi^{H}_{S}$ is a solution of the Schr\"o\-ding\-er equation. Let us suppose that
it is the homogeneous wave function defined in ($\ref{eq:3.6}$). 
The wave functions $\Psi^{\prime}_{H}$ are small perturbations.

Because the wave function $\Psi^{H}_{S}$ is a solution of the Schr\"o\-ding\-er equation,
the weight ($\ref{eq:2.10}$) is given by
\be
Weight\left[\Psi\right] = \exp{ \left[\frac{i}{\alpha \hbar} \int_{t_{1}}^{t_{2}} 
\prod_{i=1}^{N} d^{3}
{\bf r}_{i} \Psi_{H}^{\prime *} 
\hat{O} \Psi^{\prime}_{H}\right]}.
\label{eq:3.20a}
\ee
Because
\be
\Psi_{H}^{\prime} << \Psi_{S}^{H},
\label{eq:3.20b}
\ee
the measure ($\ref{eq:3.5}$) can be replaced by
\be
\prod_{t} \prod_{n=1}^{M} \frac{d \Psi_{H}^{\prime} d \Psi_{H}^{\prime *}}
{\left| \Psi_{S}^{H}(t,x_{n} )\right|^{4}}.
\label{eq.3.21}
\ee
The wave function $\Psi^{H}_{S}$ is some fixed wave function. Hence
\be
d(\Psi_{S}^{H} + \Psi_{H}^{\prime}) = d \Psi_{H}^{\prime}.
\label{eq:3.22}
\ee
Now we perform the following variable change
\be
\frac{1}{\sqrt{\alpha}} \Psi_{H}^{\prime} \longrightarrow \tilde{\Psi}
\label{eq:3.23}
\ee
The gaussian integration over $\Psi^{\prime}_{H}$ will be proportional to
\be
\prod_{t} \left( \sqrt{\alpha}\right)^{2 M}.
\label{eq:3.24}
\ee
If $\Psi^{H}_{S}$ is an homogeneous wave function, then the path integral ($\ref{eq:2.3}$) 
is proportional to
\be
\prod_{t} \left( \sqrt{\alpha}\right)^{2 M} ( a M)^{2 M}.
\label{eq:3.25}
\ee

Now let consider the set of wave functions
\be
\Psi=\Psi_{NS}^{I} + \Psi_{I}^{\prime}.
\label{eq:3.26}
\ee
Where $\Psi^{I}_{NS}$ is the inhomogeneous wave function ($\ref{eq:3.9}$). Let us suppose 
that $\Psi^{I}_{NS}$ is not solution of the Schr\"o\-ding\-er equation.
Wave functions $\Psi^{\prime}_{I}$ are small perturbations around the fixed wave function $\Psi^{I}_{NS}$.

Because $\Psi^{I}_{NS}$ is not solution of the Schr\"o\-ding\-er equation, the weight is now given by
\be
Weight\left[\Psi\right] = \exp{ \left[\frac{i}{\alpha \hbar} \int_{t_{1}}^{t_{2}} 
\prod_{i=1}^{N} d^{3}
{\bf r}_{i} \Psi_{I}^{\prime} 
\hat{O} \Psi^{I}_{NS} + c. c. + \cdots \right]}.
\label{eq:3.27bis}
\ee
Therefore we can replace the weight by 
\be
Weight\left[\Psi\right] \approx \exp{ \left[\frac{i}{\alpha \hbar} \int_{t_{1}}^{t_{2}} 
\prod_{i=1}^{N} d^{3}
{\bf r}_{i} \Psi_{I}^{\prime} 
\hat{O} \Psi^{I}_{NS} + c.c. \right]}.
\label{eq:3.27}
\ee
The measure is now given by
\be
\prod_{t} \prod_{n=1}^{M} \frac{d \Psi_{I}^{\prime} d \Psi_{I}^{\prime *}}
{\left| \Psi_{NS}^{I}(t,x_{n}) \right|^{4}}.
\label{eq.3.28}
\ee
The weight ($\ref{eq:3.27}$) depends on $\Psi_{I}^{\prime} / \alpha$. Hence, we perform the variable change
\be
\frac{1}{\alpha} \Psi_{I}^{\prime} \longrightarrow \tilde{\Psi}.
\label{eq:3.23bis}
\ee
Hence the path integral is now proportional to 
\be
\prod_{t} \alpha^{2} \left[ aB^{2} \left(1-aB^{2}\right)^{M-1} M \right]^{-2}.
\label{eq:3.29}
\ee
Hence if $\alpha $
\be
\alpha < \left(1 - aB^{2} \right)^{2}
\label{eq:3.30}
\ee
then the leading contribution to correlators ($\ref{eq:2.2}$) are given by wave functions around solutions of the Schr\"o\-ding\-er equation.

Now we are going to study the meaning of $\alpha $ and the condition ($\ref{eq:3.30}$).
Let us suppose that the probability amplitude
\be
\left|\Psi(t;{\bf r}_{1}, \cdots, {\bf r}_{N})\right|^{2} d^{3}{\bf r}_{1} \cdots d^{3}{\bf r}_{N}
\label{eq:3.31}
\ee
has discrete values. In other words, let us suppose that
\be
\left|\Psi(t;{\bf r}_{1}, \cdots, {\bf r}_{N})\right|^{2} d^{3}{\bf r}_{1} \cdots d^{3}{\bf r}_{N}
= \frac{n}{K} \, \, \, n=0,\cdots,K
\label{eq:3.32}
\ee
where $K$ is a very large integer.

If $K$ is very large then the discrete nature of the probability density has not experimental consequences.
On the other hand, the smallest non zero value of $(1-aB^{2})$ is $1/K$.
Hence if $\alpha $ is less than $1/K^{2}$, then the leading contribution to 
correlator ($\ref{eq:2.2}$) 
are given by solutions of the Schr\"o\-ding\-er equation.

Therefore the meaning of the adimensional constant $\alpha$ is the following.
It must be less than the amount of probability density that can be distinguished in modern experiments.
Hence if $\alpha$ is small and there are not local solutions of the wave equation then the 
leading contributions to correlators are given by solutions of the Schr\"o\-ding\-er equation and 
our approach is equivalent to the Schr\"o\-ding\-er equation.
But this is not true if there are local solutions of the Schr\"o\-ding\-er equation.  

Now we are going to study the differences between microscopic and macroscopic systems. 
But first of all, we must define the concept of macroscopic system more carefully.
Macroscopic systems are not just systems with a very large number of particles.
If the particles are free, then the number of particles is not important.
Macroscopic systems have solutions of the Schr\"o\-ding\-er equation that evolve without dispersion.
Or, in other words, the Schr\"o\-ding\-er equation has solutions that are local wave functions.
For microscopic systems solutions of the Schr\"o\-ding\-er equation spread over all space. 
For microscopic systems solutions of the Schr\"o\-ding\-er equation can be local only for short period of time.
Hence we have shown that for microscopic systems the leading contributions to correlators are 
given by solutions of the Schr\"o\-ding\-er equation.
 
Let us consider a macroscopic system. Let us suppose that the initial wave function is non local.
Hence the corresponding solution of the Schr\"o\-ding\-er equation is non local. Let us suppose 
that the wave function $\Psi_{H}$ defined in ($\ref{eq:3.6}$) is a solution of the Schr\"o\-ding\-er 
equation. Hence the contribution of the set of wave functions around the homogeneous solution is proportional 
to ($\ref{eq:3.25}$).

For macroscopic systems there are local solutions of the Schr\"o\-ding\-er equation. But if the initial wave 
function is non local then the wave function is non local during its temporal evolution. But let us suppose
that at the initial time the wave function collapse to some local wave function. This means that for a short 
time interval the wave function is not a solution of Schr\"o\-ding\-er equation. Hence the contribution to correlators 
of this wave function is very small during a short period of time. But after the collapse the wave function is
a local solution of the Schr\"o\-ding\-er equation. Therefore after the collapse the contribution to 
correlators of this wave function is larger than the contribution of the non local solution of the Schr\"o\-ding\-er equation.

Hence let us consider the following set of wave functions
\be
\Psi=\Psi_{I}^{S} + \Psi^{\prime},
\label{eq:3.33}
\ee
where $\Psi^{S}_{I}$ is the following fixed wave function.
At the initial time $\Psi^{S}_{I}$ is just $\Psi_{1} (t_{1})$.
In the time interval $(t_{1}, t_{1} + \Delta t)$, $\Psi^{S}_{I}$ collapse very fast to some local wave function.
In the interval $(t_{1} + \Delta t, t_{2})$, $\Psi^{S}_{I}$ evolve according to the Schr\"o\-ding\-er equation.

Because the system is macroscopic $\Psi^{S}_{I}$ is a local solution of the 
Schr\"o\-ding\-er equation between $(t_{1} + \Delta t, t_{2})$.
Hence in the interval $(t_{1}, t_{1} + \Delta t)$, wave functions ($\ref{eq:3.19}$) yield 
the leading contribution to correlators because
$\Psi^{H}_{S}$ is a solution of the Schr\"o\-ding\-er equation.
But in the interval $(t_{1} + \Delta t, t_{2})$, both $\Psi^{H}_{S}$ and $\Psi^{S}_{I}$ are solutions but $\Psi^{S}_{I}$ is local and $\Psi^{H}_{S}$ is non local.

Hence in the interval $(t_{1} + \Delta t, t_{2})$, wave functions ($\ref{eq:3.33}$) 
yield the leading contribution to correlators.
Hence if $t_{2}$ is large enough then the leading contribution to correlators are given 
by ($\ref{eq:3.33}$) for macroscopic systems.

The physical interpretation of $\Psi^{S}_{I}$ is very interesting. It represents a 
non local wave function at the initial time. It 
collapse very fast to some local wave function.
It evolves after the collapse according to the Schr\"o\-ding\-er equation. Hence non local solutions 
of the the Schr\"o\-ding\-er equation are eliminated in macroscopic systems. 

Let us remark that in macroscopic systems the Schr\"o\-ding\-er equation also work.
But we must perform a truncation on the set of physical states.
In other approaches\cite{4,5} the Schr\"o\-ding\-er equation is changed for macroscopic system.
Our approach work in a similar way than orthodox Quantum Mechanics.
But in orthodox Quantum Mechanics the truncation over the initial states is performed at hand.
In our approach the initial collapse to local wave functions is a consequence of the fundamental postulates.

In other approaches\cite{4,5} the initial collapse depends on something present at the 
moment of collapse. For instance the number of particles\cite{5}.
In our approach the collapse seem to depend on something that holds in the future.
For instance, let us consider the correlator
\be
\left\langle \Psi_{1}(t_{1}) | \Psi_{2}(t_{1} + \Delta t)\right\rangle
\label{eq:3.34}
\ee
for macroscopic systems.
In this case the leading contributions to ($\ref{eq:3.34}$) are given by solutions of the 
Schr\"o\-ding\-er equation. We have shown that non local solutions of the Schr\"o\-ding\-er equation
yield the leading contribution to correlators in the interval $(t_{1},t_{1}+\Delta t)$ 
but local solutions yield the leading contribution in the interval $(t_{1}+\Delta t,t_{2})$.
If time $t_{2}$ is near $t_{1}$ then the leading contribution to correlators are also given 
by solutions of the Schr\"o\-ding\-er equation without collapse. 
Hence, if after the initial collapse the system evolves without dispersion for a long time,
then the initial collapse holds, otherwise there is not initial collapse. 

This seems a contradiction. The collapse depend on the future evolution of the wave function. 
But the fundamental principle ($\ref{eq:2.2}$) that has replaced the Schr\"o\-ding\-er equation,
is not a temporal evolution equation.
Hence we do not need to use the concept of temporal evolution in our physical interpretation.

The fundamental law ($\ref{eq:2.3}$) is an abstract mathematical principle.
Correlators ($\ref{eq:2.2}$) are not probability amplitudes between states defined at different times.
We postulate that physical systems are defined by all correlators ($\ref{eq:2.2}$) computed for all wave functions and
for all abstract parameters $t_{1}$ and $t_{2}$.
Hence the physical objects are only the correlators ($\ref{eq:2.2}$).
But for some systems we can look for physical interpretations compatible with 
the usual concepts of orthodox Quantum Mechanics.
But sometimes these physical interpretations are not possible.

For microscopic systems the leading contribution to correlators are given by solutions of the Schr\"o\-ding\-er 
equation ($\ref{eq:3.19}$).
Solutions of the Schr\"o\-ding\-er equation are non local in these cases.
Hence correlators ($\ref{eq:2.2}$) are large if the wave 
functions $\Psi_{1}(t_{1})$ and $\Psi_{2}(t_{2})$ are connected 
by solutions of the Schr\"o\-ding\-er equation.

Let us remark that correlators are invariant under temporal inversion.
\be
\left\langle \Psi_{1}(t_{1}) \mid  \Psi_{2} (t_{2})  \right\rangle = -
\left\langle \Psi_{2}(t_{2}) \mid  \Psi_{1} (t_{1})  \right\rangle.
\label{eq:3.35}
\ee
For microscopic systems this is not important because the Schr\"o\-ding\-er equation is also 
invariant under temporal inversion.
But for macroscopic systems the initial collapse is an irreversible process. There is not contradiction
but we cannot interpret $\Psi_{1}$ and $\Psi_{2}$ as initial or final states. 
Let us call $\mid L \rangle $ to local wave functions and $\mid NL \rangle$ to non local wave functions.
It is trivial that
\be
\left| \left\langle NL(t_{1}) \mid  L (t_{2})  \right \rangle \right| = 
\left| \left\langle L(t_{2}) \mid  NL (t_{1}) \right \rangle \right| .
\label{eq:3.36}
\ee
But this equality does not mean that the inverse process to collapse is possible.
For macroscopic systems we have shown that
\bea
\left | \left\langle NL(t_{1}) \mid  L (t_{2})  \right\rangle \right |& >> &
\left | \left\langle NL(t_{1}) \mid  NL (t_{1}) \right\rangle \right | \nonumber \\
\left | \left\langle NL(t_{1}) \mid  NL (t_{2})  \right\rangle \right |& << &
\left | \left\langle L(t_{1}) \mid  NL (t_{2}) \right\rangle \right | .
\label{eq:3.37}
\eea
Let us suppose that $t_{1}$ is the initial time. Let us suppose that the state at $t_{1}$ is 
fixed and non local.
The first equation of ($\ref{eq:3.37}$) tell us that the larger correlators for 
the fixed initial state correspond to local final states.

Let us suppose that $t_{2}$ is the initial time. Let us suppose that the state at $t_{2}$ is 
fixed and non local.
The second equation of ($\ref{eq:3.37}$) tell us that the larger correlator for fixed 
initial state correspond to local final states.

Let us consider the set of all correlators ($\ref{eq:2.2}$).
Now let us consider the subset of correlators where the wave 
function $\Psi_{1}(t_{1})$ or $\Psi_{2}(t_{2})$ is fixed.
This fixed wave function can be interpreted as the initial wave function.
Final wave functions are given by the large correlators.
For microscopic systems there is only one final wave function given by the Schr\"o\-ding\-er equation.
But for macroscopic systems there are several final states. These states are local. Hence the 
same physical law ($\ref{eq:2.3}$) can describe irreversible and reversible processes.

Even though we have used the name correlator and the symbol $\langle \mid \rangle $ for the 
object ($\ref{eq:2.3}$), the correlators defined in this paper are very different from 
the correlators in orthodox Quantum Mechanics.
Let us consider a macroscopic system. Let us suppose that the initial state is $\Psi_{1}(t_{1})$.
Let us suppose that $\Psi^{A}_{2}$ and $\Psi^{B}_{2}$ are two local wave functions.
Let us suppose that the linear combination $\Psi^{A}_{2}+ \Psi^{B}_{2}$ is non local.
Then we have shown that
\bea
\left | \left\langle \Psi_{1}(t_{1}) \mid  \Psi_{2}^{A}(t_{2}) + \Psi_{2}^{B}(t_{2})  \right\rangle \right |& << &
\left | \left\langle \Psi_{1}(t_{1}) \mid  \Psi_{2}^{A}(t_{2}) \right\rangle \right | \nonumber \\
\left | \left\langle \Psi_{1}(t_{1}) \mid  \Psi_{2}^{A}(t_{2}) + \Psi_{2}^{B}(t_{2})  \right\rangle \right |& << &
\left | \left\langle \Psi_{1}(t_{1}) \mid  \Psi_{2}^{B}(t_{2}) \right\rangle \right | \nonumber \\
\label{eq:3.38}
\eea
Hence the correlator defined by ($\ref{eq:2.3}$) is not a scalar product. Actually 
\be
\left\langle \Psi_{1} \mid  \Psi_{2}^{A} + \Psi_{2}^{B}  \right\rangle \neq 
\left\langle \Psi_{1} \mid  \Psi_{2}^{A}  \right\rangle +
\left\langle \Psi_{1} \mid  \Psi_{2}^{B} \right\rangle
\label{eq:3.38bis}
\ee
This result is very important. In orthodox Quantum Mechanics we are 
very familiar with the superposition principle. Actually it is considered 
a fundamental principle. The fundamental mathematical operations in orthodox 
Quantum Mechanics are the Schr\"o\-ding\-er equation and the scalar product. Both are linear. But for macroscopic systems
the superposition principle does not work. Therefore it is very important that 
correlators defined in this paper are not necessarily linear objects. 

We have shown that for macroscopic systems the correlators 
\be
\langle \Psi_{1}(t_{1})| \Psi_{2}(t_{1}+\Delta t)\rangle
\label{eq:3.39}
\ee
and
\be 
\langle \Psi_{1}(t_{1})| \Psi_{2}(t_{2})\rangle 
\label{eq:3.40}
\ee
have no relation between them.
The leading contribution to ($\ref{eq:3.39}$) 
are given by solutions of the Schr\"o\-ding\-er equation.
But the leading contributions to ($\ref{eq:3.40}$)
are given by wave functions that collapse at the initial time
to some local wave function.
In general, given the correlators
\bea 
\langle \Psi_{1}(t_{1})| \Psi_{2}(t_{2})\rangle \nonumber \\ 
\langle \Psi_{1}(t_{1})| \Psi_{3}(t_{3})\rangle \nonumber \\ 
\eea
with $t_{1}<t_{2}<t_{3}$,
there is not relation between the physical interpretations of these correlators.
In general $\Psi_{2}(t_{2})$, is not an intermediate state between $\Psi_{1}(t_{1})$ and $\Psi_{3}(t_{3})$.
They are two different correlators.
\newpage
{\bf 4.}

In this section we are going to study the interaction between a microscopic system and a macroscopic system.
In the above section we showed how the new dynamical principle ($\ref{eq:2.3}$) explains the behaviour of macroscopic systems.
Now we are going to study the collapse of wave functions of microscopic systems when they interact with measurement devices.

Let us consider an electron that interact with a Wilson chamber.
This old problem was studied by Mott\cite{6}. Mott studied the full Schr\"o\-ding\-er equation of the particle and the Wilson chamber.
He showed that there are solutions of the full Schr\"o\-ding\-er equation that describe 
excited particles of the Wilson chamber in a straight line.

Hence there are solutions of the full Schr\"o\-ding\-er equation that describe the tracks that the particle leave in the Wilson chamber.
But orthodox Quantum Mechanics does not explain why these are the only possible solutions of the Schr\"o\-ding\-er equation.

We have shown in a qualitative way that if there are local solution of the Schr\"o\-ding\-er equation 
then non local solutions collapse to these local wave functions.

Let us consider a microscopic system interacting with a macroscopic system.
The total wave function is
\be
\Psi(t;{\bf r};{\bf r}_{1},\cdots,{\bf r}_{N})
\label{eq:4.1}
\ee
where the parameter ${\bf r}$ 
will be interpreted as the coordinate of the microscopic system.

Let us suppose that the particle is far away from the macroscopic systems at the initial time $t_{0}$.
Hence in $t_{0}$, there is not interaction between the microscopic and the macroscopic systems.
At $t_{1}$ the particle is near the macroscopic system.
Hence from $t_{1}$ the interaction between the particle and the macroscopic system starts.
Let us suppose that from $t_{1}$ there are local solutions of the Schr\"o\-ding\-er equation.
In other words, from $t_{1}$ the wave functions are defined around fixed positions of the 
variables $\{ {\bf r}; {\bf r}_{i} \}$ 

Let us suppose that in the initial time $t_{0}$ the wave function is
\be
\Psi_{0}(t_{0}) = \Psi(t_{0},{\bf r}) \Psi(t_{0}, {\bf r}_{1}, \cdots, {\bf r}_{N} ) .
\label{eq:4.2}
\ee
Now we are going to perform the calculation of correlators
\be
\left \langle \Psi_{0}(t_{0}) \mid \Psi_{2}(t_{2}) \right \rangle
\label{eq:4.3}
\ee
where $t_{2}> t_{1}$.
We are going to look for the wave functions $\Psi_{2}(t_{2})$ that make the correlators as 
large as possible.

We have shown in the above sections that the leading contributions 
to correlators ($\ref{eq:4.3}$) are given by
wave functions around some fixed wave function
\be
\Psi = \bar{\Psi} + \delta \Psi,
\label{eq:4.4}
\ee
where $\delta \Psi $ is very small and $\bar{\Psi}$ is a fixed wave function defined as follows.
From $t_{0}$ to $t_{1}$, $\bar{\Psi}$ is given by 
\be
\bar{\Psi}(t) = \Psi(t,{\bf r}) \Psi(t, {\bf r}_{1}, \cdots, {\bf r}_{N} ),
\label{eq:4.5}
\ee
where $\Psi(t,{\bf r})$
is a solution of the Schr\"o\-ding\-er equation of the microscopic system without the macroscopic system.
From $t_{1}$ to $t_{1} + \Delta t$, the wave function collapse to some local solution of the full Schr\"o\-ding\-er equation.
From $t_{1} + \Delta t$, the wave function evolve according to the full Schr\"o\-ding\-er equation.

The wave function $\bar{\Psi}$
vary very fast between $t_{1}$ and $t_{1} + \Delta t$.
Hence between $t_{1}$ and $t_{1} + \Delta t$, the weight can be replaced by
\be
\exp{ \left [ \frac{i}{\alpha \hbar} \int_{t_{1}}^{t_{1}+\Delta t} dt 
\prod_{i=1}^{N} d^{3} {\bf r}_{i} d^{3} {\bf r} 
\Psi^{*} \left ( -i \hbar \frac{\partial}{\partial t} .
\Psi \right ) \right ] }
\label{eq:4.6}
\ee

Now let us discretize the coordinates $\{ {\bf r} ; {\bf r}_{i} \}$
and the time.
The correlators are proportional to
\bea
\int \prod_{i=1}^{N} \prod_{{\bf r}_{i}} \prod_{{\bf r}} 
\frac{d \Psi^{*} ( t_{1} + \Delta t ; {\bf r} ; {\bf r}_{1} , \cdots , {\bf r}_{N} ) 
d \Psi ( t_{1} + \Delta t ; {\bf r} ; {\bf r}_{1} , \cdots , {\bf r}_{N} ) }
{\left | \Psi ( t_{1} + \Delta t ; {\bf r} ; {\bf r}_{1} , \cdots , {\bf r}_{N} )
\right |^{4} } \nonumber \\
\exp{\left [ \frac{i}{\alpha \hbar} \prod_{i=1}^{N} \sum_{{\bf r}_{i}} 
\sum_{{\bf r}} 
\left [ \Psi^{*}(t_{1}) \left [ \Psi(t_{1} + \Delta t ) - \Psi(t_{1}) \right ] + c.c 
\right ]  \right ]} .
\label{eq:4.7}
\eea
We have replaced partial derivatives by finite differences. 
At $t_{1}$ the wave function is defined over all space.
But at $t_{1} + \Delta t$, the wave function is localized around some fixed 
positions $\{ {\bf r}^{*} ; {\bf r}^{*}_{1}, \cdots , {\bf r}^{*}_{N} \}$.
Therefore the correlators are proportional to
\bea
\int 
\frac{d \Psi^{*} ( t_{1} + \Delta t ; {\bf r}^{*} ; {\bf r}_{1}^{*} , \cdots , {\bf r}_{N}^{*} ) 
d \Psi ( t_{1} + \Delta t ; {\bf r}^{*} ; {\bf r}_{1}^{*} , \cdots , {\bf r}_{N}^{*} ) }
{\left | \Psi ( t_{1} + \Delta t ; {\bf r}^{*} ; {\bf r}_{1}^{*} , \cdots , {\bf r}_{N}^{*} )
\right |^{4} } \nonumber \\
\exp{\left [ \frac{i}{\alpha \hbar}
\Psi^{*}(t_{1}; {\bf r}^{*} ; {\bf r}_{1}^{*} , \cdots , {\bf r}_{N}^{*}) 
\Psi(t_{1} + \Delta t; {\bf r}^{*} ; {\bf r}_{1}^{*} , \cdots , {\bf r}_{N}^{*} ) 
+ c.c \right ]} .
\label{eq:4.8}
\eea
Because the wave functions are normalized
\be
\left | \Psi(t_{1} + \Delta t ; {\bf r}^{*} ; 
{\bf r}_{1}^{*} , \cdots , {\bf r}_{N}^{*})
\right |^{2} \leq  1 .
\label{eq:4.9}
\ee

Now let us perform the following variable change
\be
\Psi(t_{1} + \Delta t; {\bf r}^{*}; {\bf r}^{*}_{1} , \cdots , {\bf r}^{*}_{N}) \longrightarrow 
\tilde{\Psi} ,
\label{eq:4.10}
\ee
where 
\be
\tilde{\Psi} = \frac{\Psi(t_{1})}{\alpha} \Psi 
( t_{1}+\Delta t; {\bf r}^{*}; {\bf r}^{*}_{1} , \cdots , {\bf r}^{*}_{N}) .
\label{eq:4.11}
\ee
Now the condition ($\ref{eq:4.9}$) becomes
\be
\left | \tilde{\Psi} \right |^{2} \leq \frac{\left |\Psi(t_{1}) \right |^{2}}{\alpha} .
\label{eq:4.12}
\ee
But $\alpha $ is very small. Actually $\alpha $ is smaller than the amount of probability density that 
can be detected in physical experiments.
Hence
\be
\frac{\left |\Psi(t_{1}) \right |^{2}}{\alpha}
\label{eq:4.13}
\ee
is very large and we can replace the condition ($\ref{eq:4.12}$) by
\be
\left | \tilde{\Psi} \right |^{2} \leq \infty .
\label{eq:4.14}
\ee
Hence the correlators are proportional to
\be
\left | 
\Psi (t_{1}; {\bf r}^{*}; {\bf r}^{*}_{1} , \cdots , {\bf r}^{*}_{N}) 
\right |^{2} 
\int \frac{d \tilde{\Psi} d \tilde{\Psi}^{*}}
{ \left | \tilde{\Psi} \right |^{4}}
\exp{\left [\frac{i}{\hbar} \tilde{\Psi} \right ]} .
\label{eq:4.15}
\ee
The above integral is not well defined.
But we have shown in a qualitative way that the correlators are proportional to
\be
\left | 
\Psi (t_{1}; {\bf r}^{*}; {\bf r}^{*}_{1} , \cdots , {\bf r}^{*}_{N}) .
\right |^{2} 
\label{eq:4.16}
\ee
On the other hand it is not important the values of the correlators but we are interested 
in the relationship between the correlators.

We have studied in a qualitative way the set of correlators
\be
\left \langle  \Psi_{0}(t_{0}) \mid \Psi_{2}(t_{2}) \right \rangle
\label{eq:4.17}
\ee
for a fixed wave functions $\Psi (t_{0})$ given by ($\ref{eq:4.2}$).

Because the particle interact with a measurement device there are local solutions of the 
Schr\"o\-ding\-er equation.
Actually this can be the definition of measurement devices.

The leading contributions to ($\ref{eq:4.17}$) have the following physical interpretations.
The wave function $\Psi (t_{0})$ evolves according to the Schr\"o\-ding\-er equation until $t_{1}$.
At $t_{1}$ the wave function collapses to some local wave function centered around the 
coordinates $({\bf r}^{*}; \{{\bf r}^{*}_{i}\})$.
From $t_{1}$ to $t_{2}$ the wave function evolves according to the full Schr\"o\-ding\-er equation.

There are several final wave functions $\Psi_{2}(t_{2})$.
We have shown in a qualitative way that correlators are proportional to the probability density of the wave functions 
just before their collapse.
The probability is evaluated at the collapse point
$\left ( {\bf r}^{*};{\bf r}_{1}^{*}, \cdots , {\bf r}_{N}^{*} \right )$
In other words correlators are proportional to
\be
\left | 
\Psi (t_{1}; {\bf r}^{*}; {\bf r}^{*}_{1} , \cdots , {\bf r}^{*}_{N}) .
\right |^{2} 
\label{eq:4.18}
\ee
There is not interactions between the particle and the measurement device in the 
time interval $(t_{0},t_{1})$.
Therefore the wave function at $t_{1}$ must be the product of a wave function that depends 
only on $(t_{1};{\bf r}^{*})$ and other wave function that depends 
on $(t_{1};{\bf r}^{*}_{1} , \cdots ,{\bf r}^{*}_{N})$.
Hence correlators are proportional to
\be
\left | 
\Psi (t_{1}; {\bf r}^{*}) .
\right |^{2} 
\label{eq:4.19}
\ee
This is just the measurement postulate.

In this approach the only mathematical operation that has sense is the correlator.
Sometimes there are a few wave functions that yield the leading contributions to correlators.
From these wave functions we can make some physical interpretations.

In this section we have shown that the leading contributions are given by wave functions
that evolve from $t_{0}$ to $t_{1}$, according to the Schr\"o\-ding\-er equation,
they collapse at $t_{1}$ and they leave a well defined track until $t_{2}$.
But if we compute
\be
\left \langle  \Psi_{0}(t_{0}) \mid \Psi_{1}(t_{1}) \right \rangle
\label{eq:4.20}
\ee
the leading contribution is given by a wave function that do not collapse at $t_{1}$.
Hence there is not relation between correlators ($\ref{eq:4.3}$) and ($\ref{eq:4.20}$).
There are not relationships between correlators defined at different times.
Hence physical interpretations of correlators at different times can be different.

Moreover we have shown that the correlator is not a scalar product.
For instance, tracks on a Wilson chamber are defined by local wave functions.
But linear combination of tracks are not defined by local wave functions.
Hence we observe only well defined tracks in a Wilson chamber.
\newpage
{\bf 5.}

In this paper we have introduced a new fundamental principle ($\ref{eq:2.3}$) instead
of the Schr\"o\-ding\-er equation.
We have showed in a qualitative way, that the new principle unify the Schr\"o\-ding\-er equation and
the measurement postulates.

The main purpose of this paper is to look for new fundamental laws that do not distinguish 
between physical systems and measurement devices. The two postulates introduced in this paper 
do not use the concept of measurement device. 

This new approach has the following problems.
The new dynamical law ($\ref{eq:2.3}$) has been defined using the path integral approach.
Unfortunately this kind of path integrals are well defined in some cases only.
Actually we cannot know if these path integrals are well defined until we perform all the calculations.
In this paper we have supposed that these path integrals are well defined and 
we have done qualitative arguments with the discrete version of these path integrals.
Hence, the rigorous definition of ($\ref{eq:2.3}$) is an open problem.

On the other hand, we have introduced a new adimensional coupling constant $\alpha $.
Moreover, the new law ($\ref{eq:2.3}$) is non relativistic. Hence, this new law cannot be a fundamental law.
It must arise from a more fundamental law.
This more fundamental law must be relativistic and it must have some dimensional coupling constant.
Therefore, the meaning of $\alpha $ is also an open problem.

We have studied the new principle in a qualitative way.
We have considered microscopic systems, macroscopic systems an microscopic systems interacting with
measurement devices.

For microscopic systems, the new principle is equivalent to the Schr\"o\-ding\-er equation.
For macroscopic systems, the new principle is also equivalent to the Schr\"o\-ding\-er equation.
But in this case, the new principle truncates the set of physical states.
For microscopic systems interacting with macroscopic systems, the new principle is equivalent to 
the measurement postulates.

We have shown these results in a qualitative way. We do not have developed them too far.
But they are very interesting. First of all, we have not need any additional postulate.
There are not additional conditions over the initial states.

The results do not arise from a particular model. They are consequences of the essential differences 
between microscopic and macroscopic systems.
For macroscopic systems, there are local solutions of the Schr\"o\-ding\-er equation.
For microscopic systems, all solutions of the Schr\"o\-ding\-er equation are non local.

The approached showed in this paper is based on two fundamental ideas. 
First of all, we have separated the fundamental physical laws and their physical interpretation.
The second idea is that the fundamental law ($\ref{eq:2.3}$) is not an evolution temporal equation.

In orthodox Quantum Mechanics the fundamental postulates include the physical interpretation
of the mathematical objects. Hence the concept of measurement device is present at the 
fundamental level. In this new approach the postulates are abstract mathematical principles. Physical interpretations 
are done a posteriori ant they depend on the particular physical 
system under study. 

The fundamental dynamic equation in orthodox Quantum Mechanics is the Schr\"o\-ding\-er equation. This is a temporal 
evolution equation. The new dynamic principle ($\ref{eq:2.3}$) is not a temporal evolution equation. In some cases we 
can interpret the correlators using single wave functions that 
evolve according to the Schr\"o\-ding\-er equation. But time evolution is not a fundamental concept in this approach. 

In the famous paper of Einstein, Podolsky and Rossen \cite{7}, Quantum Mechanics is criticized 
by considering intermediate times between measurement. Bohr answered to these critics establishing 
that we can only done physical interpretation of full experiments: from the preparation of the systems to 
their final interactions with measurement devices. But the Schr\"o\-ding\-er equation defines all the 
intermediate states. But in order to make the physical interpretation of intermediate states we must perform 
intermediate measurement, in other words we must perform a different physical experiment. In orthodox Quantum 
Mechanics we can talk only about results of measurements. Waves functions between measurements have no sense.
If we perform intermediate measurements we have a different experiment and its physical interpretation 
is also different.  

In our approach the only mathematical fundamental operation is the correlator
\be 
\left\langle \Psi_{0}(t_{0}) | \Psi_{2}(t_{2}) \right\rangle 
\label{eq:5.1}
\ee
Therefore the values of wave functions at intermediate times between $t_{0}$ and $t_{2}$ have no sense. Sometimes the 
leading contribution to correlators is given by one wave function. In this case we can talk about intermediate states. 

Now let us consider the correlator 
\be 
\left\langle \Psi_{0}(t_{0}) | \Psi_{1}(t_{1}) \right\rangle 
\label{eq:5.2}
\ee
where $t_{1}$ is an intermediate time between $t_{0}$ and $t_{2}$. Sometimes the leading contributions to both 
correlators ($\ref{eq:5.1}$) and ($\ref{eq:5.2}$) are given by the same wave function. Hence $\Psi_{1}(t_{1})$ 
is the intermediate state between $t_{0}$ and $t_{2}$. But sometimes the leading contributions to both 
correlators are given by different wave functions. In these cases the physical interpretation of ($\ref{eq:5.1}$) 
and ($\ref{eq:5.2}$) are different.

Hence only correlators have sense. Wave functions at intermediate times have no sense. Correlators at different 
times are different objects with different physical interpretation. This approach is very similar to 
the answer of Bohr to \cite{7}. But in this approach we do not use the concept of measurement device. 

In this approach time is not a fundamental concept. It is part of the physical interpretation of correlators. Let 
us remark that the new principle ($\ref{eq:2.3}$) unify the smooth reversible
evolution of the Schr\"o\-ding\-er equation and the sudden irreversible collapse of the wave 
function into a single postulate. Therefore it is natural that time plays a peculiar role in our 
approach.

But if time is not a fundamental concept we must face with the introduction of relativity in this formalism. 
The new principle is analogous to the second quantization of the Schr\"o\-ding\-er equation. 
Hence in order to introduce relativistic principles is natural to go to third quantization. Our first 
postulate is non relativistic. The natural generalization of the non relativistic wave functions
\be
\Psi(t;{\bf r}_{1}, \cdots, {\bf r}_{N})
\label{eq:5.3}
\ee
are functionals of fields
\be
\Psi\left[ A_{\mu}, \Psi_{\alpha}, \Phi, \cdots \right] .
\label{eq:5.4}
\ee
But it is not clear the role of relativistic invariance in this approach. Perhaps Lo\-rentz covariance is not a 
fundamental principle. Perhaps Lorentz covariance arises from more abstract mathematical principles.

We have introduced only two postulates. We have no talk about operators. The definition of operator 
in orthodox Quantum Mechanics is related to the measurement postulates. Therefore we suppose that operators
are not fundamental concepts. But they arise when we perform the physical interpretation.

\end{document}